%% file: main.tex
\documentclass[a4paper]{article}
\usepackage[utf8]{inputenc}
\usepackage{INTERSPEECH2020}
\usepackage{cite}
\usepackage{hyperref}
\usepackage{xcolor}
\usepackage{amsmath}

\title{Aphasic Speech Recognition using a Mixture of Speech Intelligibility Experts}
\name{Matthew Perez, Zakaria Aldeneh, Emily Mower Provost}
\address{
  University of Michigan, Ann Arbor}
\email{mkperez@umich.edu, aldeneh@umich.edu,emkprovost@umich.edu}

\begin{document}

\maketitle




\begin{abstract}
Robust speech recognition is a key prerequisite for semantic feature extraction in automatic aphasic speech analysis.
However, standard one-size-fits-all automatic speech recognition models perform poorly when applied to aphasic speech.
One reason for this is the wide range of speech intelligibility due to different levels of severity (i.e., higher severity lends itself to less intelligible speech).
To address this, we propose a novel acoustic model based on a mixture of experts (MoE), which handles the varying intelligibility stages present in aphasic speech by explicitly defining severity-based experts.
At test time, the contribution of each expert is decided by estimating speech intelligibility with a speech intelligibility detector (SID).
We show that our proposed approach significantly reduces phone error rates across all severity stages in aphasic speech compared to a baseline approach that does not incorporate severity information into the modeling process.
\end{abstract}
\noindent\textbf{Index Terms}: disordered speech recognition, aphasia, speech intelligibility, mixture of experts, automatic speech recognition

\section{Introduction}
Aphasia is an acquired language disorder that manifests itself in speech~\cite{danly1982speech,ash2010speech}.
In the U.S., approximately two million people are living with aphasia and more than 180,000 acquire it every year due to brain injury, most commonly from a stroke \cite{AphasiaAssoc}. 
Aphasic speech can be difficult to understand, and as a result, persons with aphasia (PWAs) are typically characterized as having low speech intelligibility. 
Previous studies have shown that speech-language therapy activities have a positive effect on the communication abilities of PWAs~\cite{basso1992prognostic,bhogal2003intensity}. However, current speech-language analyses and speech therapy activities require in-clinic visits with a trained speech-language pathologist, which is both costly and time-consuming. Speech-based technology provides an attractive avenue for assisting medical professionals due to its low cost, high accessibility, and promising results for automating processes that relate to the assistance and analysis of those with disordered speech~\cite{christensen2014automatic,le2016automatic,le2018automatic,fraser2013automatic,le2016improving,perez2018classification}. Many of these automatic analyses rely on accurate speech transcriptions, which motivates the need for better automatic speech recognition (ASR). 



There are several challenges associated with building robust ASR systems for aphasic speech, which include low speech intelligibility~\cite{mengistu2011comparing}, high (inter- and intra-) speaker variability~\cite{hawley2007speech,mustafa2015exploring}, and relatively limited data~\cite{christensen2012comparative}. 
Some artifacts of low speech intelligibility include halting speech, use of jargon, as well as various phone-level and word-level substitutions (paraphasias). 
These speech impairments can also be compounded by co-occurring motor control disorders such as apraxia of speech and dysarthria~\cite{jordan2006disorders}.
Additionally, in regards to machine learning efforts, it's difficult to apply large, monolithic networks that traditionally take advantage of big data due to the low-resource domain of aphasic speech.

Prior works have focused on improving ASR for disordered speech by addressing the issue of speaker variability through speaker-selective model adaptation~\cite{christensen2012comparative} and speaker embeddings~\cite{le2016improving}. 
In modeling speaker variability, it is likely that these models also capture speech intelligibility, as it is an underlying feature of speaker variability.
Therefore, we believe that the success of these approaches demonstrates the potential for modeling speech intelligibility for disordered speech recognition.

In this work, we present a novel approach that explicitly models speech intelligibility in the acoustic model of an ASR system to provide more robust aphasic speech recognition.
To accomplish this, we use an acoustic model architecture designed to handle the wide range of speech intelligibility present in aphasic speech.
Specifically, we use a Mixture of Experts (MoE), deep neural network (DNN) acoustic model where each expert in the model focuses on specific severity classes defined using the speaker Aphasia Quotient (AQ) severity metric\cite{kertesz1974aphasia}.
We introduce a \emph{speech intelligibility detector} (SID), trained to detect severity levels of a given speech frame, to automatically estimate the contribution of each expert in our proposed acoustic model. 
Our results demonstrate that the proposed MoE acoustic model improves phone recognition performance over a traditional one-size-fits-all acoustic model across all speaker severity levels.





\section{Related Works}
Disordered speech recognition is an active area of study due to its important role in automatic speech-based analyses and potential impact on remote health monitoring applications.
The disordered speech recognition domain shares many challenges with the accented speech recognition domain. Several approaches such as joint modeling, multitask-learning, and MoEs have been proposed in the field of accented speech recognition to address these challenges ~\cite{jain2018improved,yang2018joint,jain2019multi}.

Le and Mower Provost improved aphasic speech recognition performance by adapting the acoustic model through appending input acoustic features with fixed-length speaker identity vectors (i-vectors)~\cite{le2016improving}.
i-vectors reduce speaker variability in an acoustic model by providing information about general speaker characteristics (e.g., pronunciation patterns)~\cite{saon2013speaker}.
Christensen et. al. showed that speaker selective training improves isolated word recognition on disordered speech~\cite{christensen2012comparative}. The authors applied maximum a posteriori (MAP) adaptation to various HMM-GMM acoustic models with a select pool of speakers. They found that speaker-independent models trained on non-dysarthric speech were inferior to models trained on dysarthric speech for the task of disordered speech recognition, highlighting the importance of acoustic similarity in disordered speech recognition.


Findings from previous research suggest that careful consideration should be taken when training acoustic models for disordered speech applications as a high variation in speech intelligibility exists among speakers.
One approach to address the high variation present in disordered speech is to train several independent acoustic models, one for each intelligibility group.
However, the main challenge with training independent models is the limited amount of data available for each group.
The proposed MoE acoustic model mitigates this challenge by allowing us to train a single network that explicitly decomposes the acoustic model into several \emph{expert} acoustic models, where each expert is focused on a specific range of speech intelligibility.
MoE based approaches in acoustic modeling have proven to be successful in accented speech recognition tasks by allowing acoustic models to handle several accents simultaneously~\cite{jain2019multi}. 
To the best of our knowledge, we are the first to explore utilizing speech intelligibility in an acoustic model for aphasic speech recognition.

\begin{figure}[t]
    \def\svgwidth{\linewidth}
    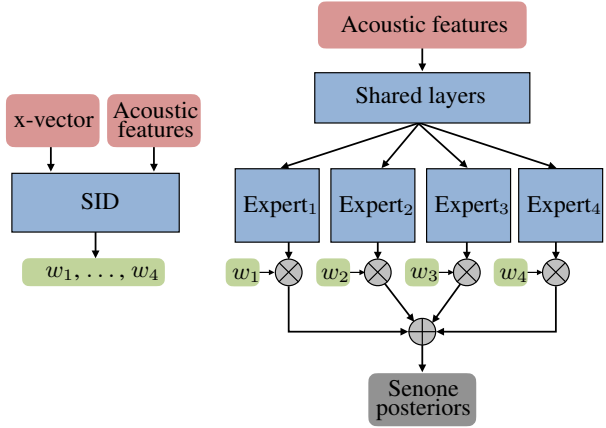
  \caption{The proposed speech intelligibility identifier (SID) and mixture of experts acoustic model. The contributions of the individual experts is determined by the outputs from the SID.}
  \vspace{-12pt}
  \label{fig:MoE}
\end{figure}

\section{Data}\label{sec_data}
AphasiaBank is a large audiovisual database composed of multiple sub-databases, which contain interactions between PWAs and clinicians\cite{macwhinney2011aphasiabank}. 
Sub-databases contain a mix of speakers with varying severity and aphasia type (i.e., Broca's, Conduction, etc.). Some sub-databases also include healthy speakers who followed a similar elicitation protocol. 
The elicitation protocol consists of free speech (i.e., personal story, describing current day), semi-structured speech (i.e., picture description tasks), and  procedural discourse (i.e. ``How would you make a peanut butter and jelly sandwich?").
In this work, we focus on English only sub-databases containing at least four speakers. 

The final dataset we use contains 25 sub-databases with 106,509 utterances and over 95 hours of speech. There are 537 speakers (237 healthy, 138 mild-aphasia, 115 moderate-aphasia, 46 severe-aphasia). 
Of these speakers, 241 are male and 210 are female, with the average age being 62.5$\pm14.5$ years.
Our metric for speech intelligibility is the WAB-R Aphasia Quotient (AQ), which is a measurement for aphasic severity~\cite{kertesz1974aphasia}. 
AQ scores $\geq 75$ are considered mild, scores of 75-50 are considered moderate, and scores $< 50$ are considered severe. 

We randomly split the dataset into training, validation, and test partitions using a 70/5/25 rule. 
Our test set is formed by randomly selecting 25\% of the data in each sub-database. 
In addition, our splits are speaker-independent and severity-stratified to ensure similar distributions of severity across the three partitions.
All recordings were down-sampled to 16kHz in this work.

\section{Method}\label{sec_method}



The architecture for the proposed MoE acoustic model is shown in Figure~\ref{fig:MoE}.
The inputs to the model are acoustic features and the outputs are context-dependent triphone state probabilities (senones).
The acoustic model consists of two parts: (1) shared initial layers used for feature extraction and (2) several expert networks each focusing on a pre-defined severity group.
We fix the number of experts to four (healthy, mild, moderate, and severe) following the speech intelligibility levels defined in Section~\ref{sec_data}.
The shared layers of the model are designed to capture broad phonetic representations applied across all severities, while the experts are designed to reduce speech intelligibility variation within their respective networks.
All hidden layers are fully-connected, consist of 1024 nodes, and use a ReLU activation function.
The shared layers of the model consist of four hidden layers and each expert is composed of two hidden layers and the final output layer applies a softmax function. Our baseline network is a fully connected DNN with 6 hidden layers.

In addition to the acoustic model, we also propose a speech intelligibility detector (SID) which predicts speech intelligibility levels, as quantified by the AQ severity metric, given the acoustic features and utterance-level speaker embeddings. 
The outputs from the SID are used as weights for computing the contribution of each expert in our acoustic model.
We use speaker embeddings extracted from a pre-trained x-vector system~\cite{snyder2018x} and apply principal component analysis (PCA) to reduce the dimensionality of these x-vectors from 512 to 32 to avoid over-fitting.
x-vectors have previously been used to estimate speech intelligibility~\cite{chandrakala2020bag} due to their ability to encode many properties about a speaker (i.e. speaking style, speaking rate, etc.)~\cite{williams2019disentangling,raj2019probing}.
The SID consists of two hidden layers and a final output layer which applies a softmax function.

The output of the acoustic model at time frame $x$ is determined using the following equation:

\begin{equation}
    \label{eq:output_oracle_SID}
    y(x)=\sum_{i=0}^N w_i(x)M_i(x)
\end{equation}
where $y(x)$ are the final senone probabilities, $w_i(x)$ is the contribution of expert $i$ as computed by the SID,  $M_i$ are the senone probabilities from expert $i$, and $N$ is the total number of experts in the model.

\begin{table}[t]
\caption{Data sharing protocols used when training the MoE acoustic models. $M_i$ represents the $i$-th expert in the model. $h$, $mi$, $mo$, and $s$ represent utterances from speakers that are Healthy, Mild, Moderate, and Severe, respectively.}
\label{tab:data_sharing3}
\centering
\begin{tabular}{l|cccc}
\toprule
\multicolumn{1}{c|}{\textbf{Setup}} & $\pmb{M_0}$ & $\pmb{M_1}$ & $\pmb{M_2}$ & $\pmb{M_3}$ \\
\midrule
$Solo$          & $h$ & $mi$       & $mo$        & $se$ \\
$Solo$+$Healthy$  & $h$ & $h\cup mi$ & $h\cup mo$  & $h\cup se$ \\
$Solo$+$Neighbor$ & $h$ & $h\cup mi$ & $mi\cup mo$ & $mo\cup se$ \\
\bottomrule
\end{tabular}
\vspace{-3pt}
\end{table}

\begin{table*}[t]
\caption{Model performance (in PER) based on severity classes. Relative PER improvement is shown in parenthesis and an * is used to indicate a PoI of \textgreater99\% over the baseline, using the method proposed in~\cite{bisani2004bootstrap}. Best performance for each setup is in \textbf{bold}.}
~\label{tab:oracleSID_sev}
\centering
\begin{tabular}{l||c|ccc}
\toprule
\multicolumn{1}{c||}{\textbf{Model}}  &\textbf{Overall}  & \textbf{Mild} & \textbf{Moderate} & \textbf{Severe} \\
\midrule
Baseline & 37.96 & 34.19 & 42.45 & 66.56  \\
$Solo$ & 40.29 (-5.8) & 36.21 (-5.6) & 44.79 (-5.2) & 77.38 (-13.9)  \\
$Solo$+$Healthy$ & 38.35 (-1.0) & 34.54 (-1.0) & 42.77 (-0.7) & 70.03 (-4.9)  \\
$Solo$+$Neighbor$ & \textbf{37.03 (+2.5)*} & \textbf{33.37 (+2.4)*} & \textbf{41.69 (+1.8)*} & \textbf{61.41 (+7.7)*}  \\
\bottomrule
\end{tabular}
\end{table*}

\section{Experimental Setup}
We use the Kaldi toolkit~\cite{povey2011kaldi} for feature extraction and HMM-GMM training and use the pytorch-kaldi toolkit~\cite{ravanelli2019pytorch} for neural network training.
The feature extraction pipeline involves the extraction of 13-dimensional mel frequency cepstral coefficient (MFCC) features, with cepstral mean and variance normalization (CMVN) spliced across 7 frames ($\pm$3).
Frames are created using a sliding window of size 25 ms with a 10 ms shift rate.
We apply Linear Discriminant Analysis (LDA) to reduce the dimension size to 40 and use Maximum Likelihood Linear Transformation (MLLT) to further decorrelate the features~\cite{gopinath1998constrained}. 
Lastly, we apply feature-space Maximum Likelihood Linear Regression (fMLLR) \cite{povey2006feature} as a means of speaker adaptation and use the resulting output as acoustic features for HMM-DNN training.
The final inputs to the DNN acoustic models are 40-dimensional fMLLR acoustic features with a context window of 11 ($\pm$5) frames, for a total feature size of 440.

We train our networks with an SGD optimizer using an initial learning rate of 0.01.
The learning rate is halved after each epoch when the performance on the validation set stagnates.
We use early stopping to reduce overfitting.
For evaluation, we use Phone Error Rate (PER), following the same approach as \cite{le2016improving}. 
To assess the statistical significance of PER improvement between our models, we use a bootstrap estimation test provided by the Kaldi \texttt{compute-wer-bootci} tool~\cite{bisani2004bootstrap}. 
This method computes \emph{probability of improvement} (PoI), where PER is computed on \begin{math}10^{4}\end{math} sampled utterances for a given model in order to estimate the spread of PER around its mean.
The difference in PER is then computed for each sample between the baseline and the proposed model being evaluated, providing an estimate of the probability of improvement.

\begin{table*}[t]
\caption{Model performance (in PER) when using a trained SID at test time. All models are trained following the $Solo+Neighbor$ approach. * indicates a PoI of \textgreater99\% over the baseline, using the method proposed in~\cite{bisani2004bootstrap}. Best performance for each setup is in \textbf{bold}.}
~\label{tab:pretrain_SID}
\centering
\begin{tabular}{l||c|ccc}
\toprule
\multicolumn{1}{c||}{\textbf{Model}} & \textbf{Overall} & \textbf{Mild} & \textbf{Moderate} & \textbf{Severe} \\
\midrule
Baseline & 37.96 & 34.19 & 42.45 & 66.56  \\


SID$_{frame}$ & 36.98 (+2.6)* & 33.20 (+2.9)*& 41.67 (+2.0)*& \textbf{62.71 (+5.8)}*  \\ 
SID$_{utt}$ & \textbf{36.87 (+2.9)}*& \textbf{33.07 (+3.3)}*& \textbf{41.64 (+2.0)}*& 62.86 (+5.6)* \\ 
\bottomrule
\end{tabular}
\end{table*}

\section{Experiments and Results}
\subsection{MoE with Oracle Speech Intelligibility Detector}
\label{sec:MoE_oracle}
In our initial set of experiments, we assume that we have access to an \emph{oracle} SID, which uses the speaker's known AQ to select the proper expert at test time. 
The goal of this section is two-fold: (1) explore the relationship between data assignment among experts and acoustic model performance during training and (2) demonstrate the effectiveness of incorporating information about speech intelligibility into an MoE acoustic model.
We experiment with several models to study the effect of data assignment among experts. 

The \emph{baseline} model that we start with has the same architecture as the model described in Section~\ref{sec_method} with the exception of only having one expert. 
The baseline is a one-size-fits-all model that is typically used for general-purpose ASR applications.
Next, we introduce a series of data assignment methods for training the MoE, outlined in Table~\ref{tab:data_sharing3}.

The first approach, $Solo$, is where each expert in the MoE acoustic model only sees data samples from one severity group.
The $Solo$ model can be thought of as the other extreme to our baseline model;
whereas the baseline model pools all data regardless of its severity group, the $Solo$ model splits data among experts based on their respective severity group.

One challenge with the $Solo$ approach, however, is that each expert in the model will see a limited amount of data. 
To compensate for this challenge, we introduce the $Solo+Healthy$ approach, which augments the data seen by each expert with healthy speech samples.
While this might improve the performance over the $Solo$ approach, we hypothesize that large variation in speech intelligibility present among healthy and aphasic speakers might hinder model performance.

We finally introduce the $Solo+Neighbor$ approach, which is used to augment the data seen by each expert while addressing the data discrepancy challenges that could arise from using the $Solo+Healthy$ approach.
In the $Solo+Neighbor$ approach, we augment the data seen by each expert by incorporating data that is similar in terms of speech intelligibility.
Specifically, each expert in our $Solo+Neighbor$ acoustic model sees both its respective data and data from the neighboring expert (i.e., the expert focusing on lower severity class).

\subsubsection{Results}
\label{sec:oracle-results}
Table~\ref{tab:oracleSID_sev} shows the model PER performance across all severity classes as well as the relative improvement and PoI over the baseline. 
We find that a MoE acoustic model, combined with an oracle SID and proper data assignment training ($Solo+Neighbor$), results in significantly lower PERs across all severity levels when compared to PERs obtained using a standard one-size-fits-all acoustic model.

First, we find that the $Solo$ approach yields higher PERs compared to those obtained from the baseline.
We believe this drop in performance is due to a lack of training data in each expert as the restrictive nature of $Solo$ only exacerbates a data scarcity problem.
As we move to a less restrictive data assignment method ($Solo+Healthy$), we can see improved performance over the $Solo$ approach as healthy data is given to each expert. 
However, in comparison to the baseline, $Solo+Healthy$ still performs worse, which we believe is due to the dissimilarity (in speech intelligibility) between healthy samples and the original $Solo$ samples in each expert.
This is confirmed by the performance of $Solo+Neighbor$, which shows that less restrictive data assignment that attempts to preserve speech intelligibility similarity yields the best results.
With $Solo+Neighbor$, we see an overall 2.5\% relative improvement over the baseline with 2.4\%, 1.8\%, and 7.7\% relative improvements for mild, moderate, and severe forms of aphasia, respectively.

The results show that when training an MoE based on speech intelligibility, being overly restrictive with expert training can compound the effects of the data scarcity problem. 
We show that this can be alleviated through non-mutually exclusive data assignment and that preserving speech intelligibility consistency in each expert when considering pooling strategies (shown by $Solo+Neighbor$) leads to better performance over a one-size-fits-all model. 
With these considerations, we are able to show that a MoE acoustic model trained using the $Solo+Neighbor$ approach improves phone recognition performance over a baseline when supported by an oracle SID that estimates speaker severity classes.




\subsection{MoE with Automatic Speech Intelligibility Detector}
Our previous experiments assumed that we had access to an oracle SID that perfectly predicts speech intelligibility.
In this section, the contributions of each expert in our acoustic model are guided by the predictions made by a neural SID model that is discriminatively trained to predict severity classes given the same features used by the acoustic model in addition to speaker embeddings extracted by a pre-trained x-vector model.

We use the $Solo+Neighbor$ data assignment approach when training our MoE, as it provided significant performance improvements over other setups in Section~\ref{sec:oracle-results}.
We experiment with two variants of the SID in order to study the effect that averaging speech intelligibility has on ASR performance.
More specifically, we investigate averaging SID predictions at the frame-level and at the utterance-level.
Utterance-level SID predictions force the contributions of the experts to be consistent across all frames from a given utterance, while frame-level SID predictions can result in different expert contributions from one frame to next.
We refer to the frame-level and utterance-level SIDs as $SID_{frame}$ and $SID_{utt}$, respectively.

\subsubsection{Results}
\label{sec:results_SID}
\begin{figure}
  \vspace{-13pt}
  \def\svgwidth{\linewidth}
  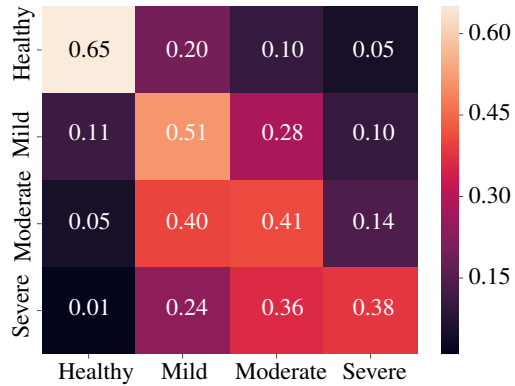
  \vspace{-14pt}
  \caption{Confusion Matrix of frame-level SID performance}
  \label{fig:SID-conf}
  \vspace{-12pt}
\end{figure}

In section~\ref{sec:MoE_oracle}, we assumed that the model had access to an oracle classifier, which was able to correctly identify the severity class of the speaker. 
Here, we train a SID using fMLLR and x-vector features to predict severity class at the frame-level. 
The performance of the SID is shown in Figure~\ref{fig:SID-conf} where we see confusion between neighboring class severities.
Although the SID is imperfect, the confusion trend between adjacent classes indicates that the SID is learning some aspects related to speech intelligibility. 
We believe that the poor SID performance is partly due to the use of AQ scores, which include other modalities beyond acoustic speech (i.e. language usage, motor control tests, family history, etc.).
In addition to this, we believe another contributing factor is that these AQ labels are speaker-level speech intelligibility scores, which assumes that speech intelligibility manifests itself at the speaker-level when, in fact, speech intelligibility may vary on a much smaller time scale. 
Without access to smaller time-scale intelligibility scores, we investigate the impact that averaging the SID output at the frame-level and utterance-level have on MoE performance.

For all experiments in this section, the neural SID outputs, at test time, are used as soft weights (probabilities) for merging the senone predictions of all experts appropriately for a given test sample. 
Table~\ref{tab:pretrain_SID} shows the PER, relative performance improvement, and PoI over the baseline, across severity classes. 

$SID_{frame}$ achieves improved results over the baseline model, across all severities, however, $SID_{utt}$ yields slightly better results with 2.9\% relative improvement overall with 3.3\%, 2.0\%, and 5.6\% relative improvements for mild, moderate, and severe respectively. 
This shows that a trained neural SID can be used to improve phone recognition in an MoE acoustic model over a traditional one-size-fits-all model.
It is also interesting to note that using an automatic SID outperformed using an oracle SID in the MoE framework.
We believe this may indicate that speaker-level speech intelligibility scores may not be as accurate as utterance-level or frame-level intelligibility scores.
Overall, these results show that despite the imperfect performance of the SID at predicting speaker-level speech intelligibility, the MoE still achieves performance improvement over a baseline model. 

\section{Conclusion}
In this work, we first show that modeling speech intelligibility in a MoE acoustic model improves phone recognition across all severities with the use of an oracle SID. 
We then demonstrate the importance of data assignment when training the MoE model.
As we move away from the use of an oracle SID, we show how a neural SID can be trained using fMLLR and x-vector features to estimate severity classes. Lastly, we show that the outputs of the SID can be used to merge expert senone predictions at test time for improved MoE performance over a one-size-fits-all acoustic model.

In section~\ref{sec:results_SID} we suggest that a potential limitation of this work is in utilizing speaker-level speech intelligibility scores. In future work, we plan to explore smaller time-scale metrics for estimating speech intelligibility and how this can be used to improve the SID and ultimately our MoE acoustic model. 

\section{Acknowledgments}
This material is based in part upon work supported by the Toyota Research Institute (``TRI'') and by the National Science Foundation Graduate Research Fellowship Program. Any opinions, findings, and conclusions or recommendations expressed in this material are those of the authors and do not necessarily reflect the views of the NSF, TRI, or any other Toyota entity.

\bibliography{bib}
\bibliographystyle{ieeetr}
\end{document}

%% file: media/figure1.eps_tex
\begingroup%
  \makeatletter%
  \providecommand\color[2][]{%
    \errmessage{(Inkscape) Color is used for the text in Inkscape, but the package 'color.sty' is not loaded}%
    \renewcommand\color[2][]{}%
  }%
  \providecommand\transparent[1]{%
    \errmessage{(Inkscape) Transparency is used (non-zero) for the text in Inkscape, but the package 'transparent.sty' is not loaded}%
    \renewcommand\transparent[1]{}%
  }%
  \providecommand\rotatebox[2]{#2}%
  \newcommand*\fsize{\dimexpr\f@size pt\relax}%
  \newcommand*\lineheight[1]{\fontsize{\fsize}{#1\fsize}\selectfont}%
  \ifx\svgwidth\undefined%
    \setlength{\unitlength}{1023.9999744bp}%
    \ifx\svgscale\undefined%
      \relax%
    \else%
      \setlength{\unitlength}{\unitlength * \real{\svgscale}}%
    \fi%
  \else%
    \setlength{\unitlength}{\svgwidth}%
  \fi%
  \global\let\svgwidth\undefined%
  \global\let\svgscale\undefined%
  \makeatother%
  \begin{picture}(1,0.75)%
    \lineheight{1}%
    \setlength\tabcolsep{0pt}%
    \put(0,0){\includegraphics[width=\unitlength]{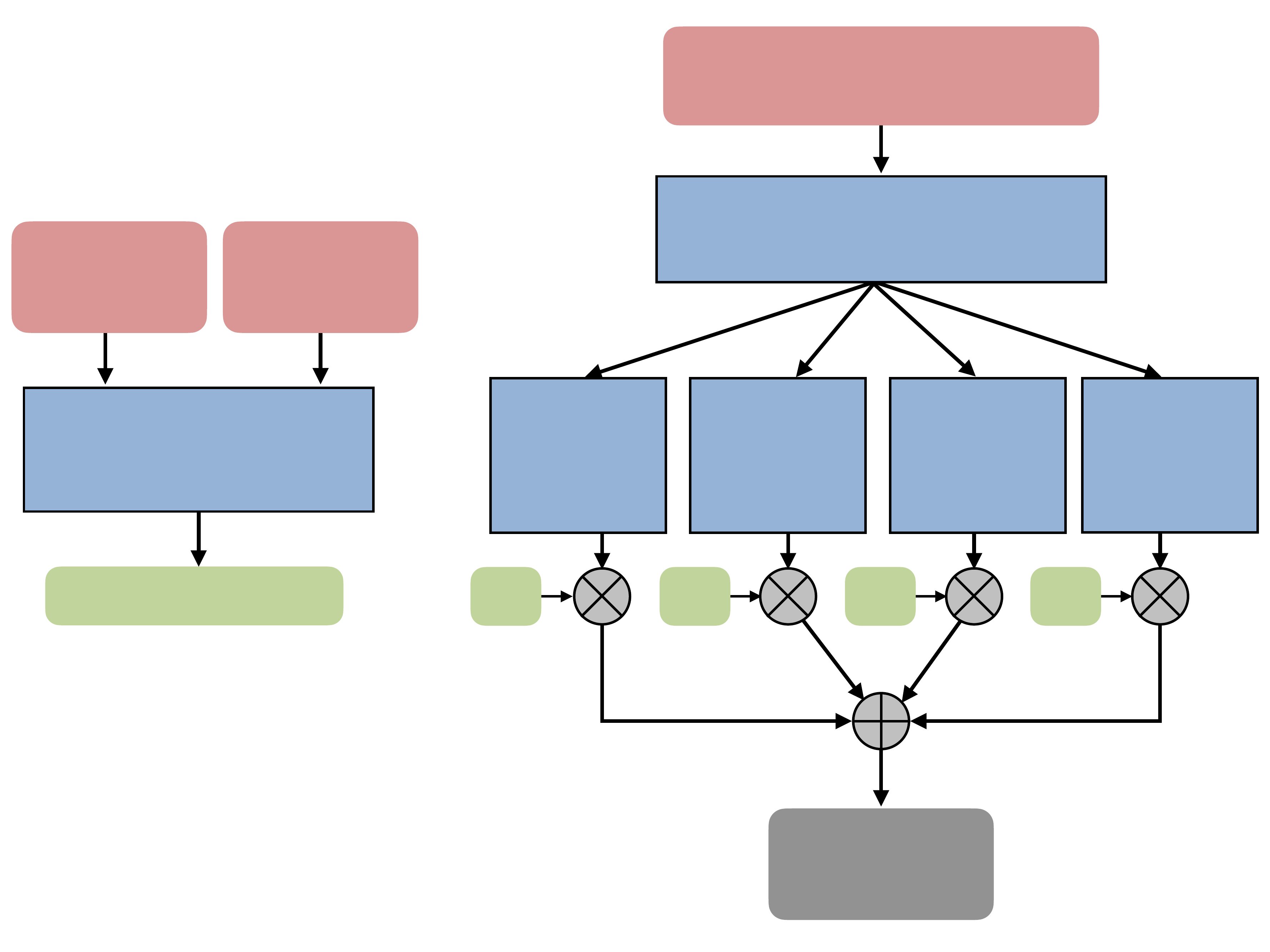}}%
    \put(0.55972822,0.67921123){\color[rgb]{0,0,0}\makebox(0,0)[lt]{\lineheight{1.25}\smash{\begin{tabular}[t]{l}Acoustic features\end{tabular}}}}%
    \put(0.58414717,0.56091768){\color[rgb]{0,0,0}\makebox(0,0)[lt]{\lineheight{1.25}\smash{\begin{tabular}[t]{l}Shared layers\end{tabular}}}}%

    \put(0.39994062,0.38047295){\color[rgb]{0,0,0}\makebox(0,0)[lt]{\lineheight{1.25}\smash{\begin{tabular}[t]{l}Expert$_1$\end{tabular}}}}%
    \put(0.55733965,0.38034824){\color[rgb]{0,0,0}\makebox(0,0)[lt]{\lineheight{1.25}\smash{\begin{tabular}[t]{l}Expert$_2$\end{tabular}}}}%
    \put(0.71473857,0.38076006){\color[rgb]{0,0,0}\makebox(0,0)[lt]{\lineheight{1.25}\smash{\begin{tabular}[t]{l}Expert$_3$\end{tabular}}}}%
    \put(0.86618652,0.38076006){\color[rgb]{0,0,0}\makebox(0,0)[lt]{\lineheight{1.25}\smash{\begin{tabular}[t]{l}Expert$_4$\end{tabular}}}}%

    \put(0.13237891,0.3847041){\color[rgb]{0,0,0}\makebox(0,0)[lt]{\lineheight{1.25}\smash{\begin{tabular}[t]{l}SID\end{tabular}}}}%

    \put(0.1850165,0.53652793){\color[rgb]{0,0,0}\makebox(0,0)[lt]{\lineheight{1.25}\smash{\begin{tabular}[t]{l}Acoustic\end{tabular}}}}%
    \put(0.1930165,0.50480527){\color[rgb]{0,0,0}\makebox(0,0)[lt]{\lineheight{1.25}\smash{\begin{tabular}[t]{l}features\end{tabular}}}}%

    \put(0.02221433,0.5206666){\color[rgb]{0,0,0}\makebox(0,0)[lt]{\lineheight{1.25}\smash{\begin{tabular}[t]{l}x-vector\end{tabular}}}}%

    \put(0.63776045,0.07595906){\color[rgb]{0,0,0}\makebox(0,0)[lt]{\lineheight{1.25}\smash{\begin{tabular}[t]{l}Senone\end{tabular}}}}%
    \put(0.61776045,0.04423641){\color[rgb]{0,0,0}\makebox(0,0)[lt]{\lineheight{1.25}\smash{\begin{tabular}[t]{l}posteriors\end{tabular}}}}%

    \put(0.0740441,0.2716373){\color[rgb]{0,0,0}\makebox(0,0)[lt]{\lineheight{1.25}\smash{\begin{tabular}[t]{l}$w_1$, $\dots$, $w_4$ \end{tabular}}}}%
    \put(0.38119756,0.26918242){\color[rgb]{0,0,0}\makebox(0,0)[lt]{\lineheight{1.25}\smash{\begin{tabular}[t]{l}$w_1$\end{tabular}}}}%
    \put(0.52765332,0.26918242){\color[rgb]{0,0,0}\makebox(0,0)[lt]{\lineheight{1.25}\smash{\begin{tabular}[t]{l}$w_2$\end{tabular}}}}%
    \put(0.67415205,0.27015498){\color[rgb]{0,0,0}\makebox(0,0)[lt]{\lineheight{1.25}\smash{\begin{tabular}[t]{l}$w_3$\end{tabular}}}}%
    \put(0.82063643,0.26918242){\color[rgb]{0,0,0}\makebox(0,0)[lt]{\lineheight{1.25}\smash{\begin{tabular}[t]{l}$w_4$\end{tabular}}}}%
  \end{picture}%
\endgroup%

%% file: media/out_xvec_fmllr_classifier_SID_norm_conf_matrix.eps_tex
\begingroup%
  \makeatletter%
  \providecommand\color[2][]{%
    \errmessage{(Inkscape) Color is used for the text in Inkscape, but the package 'color.sty' is not loaded}%
    \renewcommand\color[2][]{}%
  }%
  \providecommand\transparent[1]{%
    \errmessage{(Inkscape) Transparency is used (non-zero) for the text in Inkscape, but the package 'transparent.sty' is not loaded}%
    \renewcommand\transparent[1]{}%
  }%
  \providecommand\rotatebox[2]{#2}%
  \newcommand*\fsize{\dimexpr\f@size pt\relax}%
  \newcommand*\lineheight[1]{\fontsize{\fsize}{#1\fsize}\selectfont}%
  \ifx\svgwidth\undefined%
    \setlength{\unitlength}{460.80098848bp}%
    \ifx\svgscale\undefined%
      \relax%
    \else%
      \setlength{\unitlength}{\unitlength * \real{\svgscale}}%
    \fi%
  \else%
    \setlength{\unitlength}{\svgwidth}%
  \fi%
  \global\let\svgwidth\undefined%
  \global\let\svgscale\undefined%
  \makeatother%
  \begin{picture}(1,0.75000271)%
    \lineheight{1}%
    \setlength\tabcolsep{0pt}%
    \put(0,0){\includegraphics[width=\unitlength]{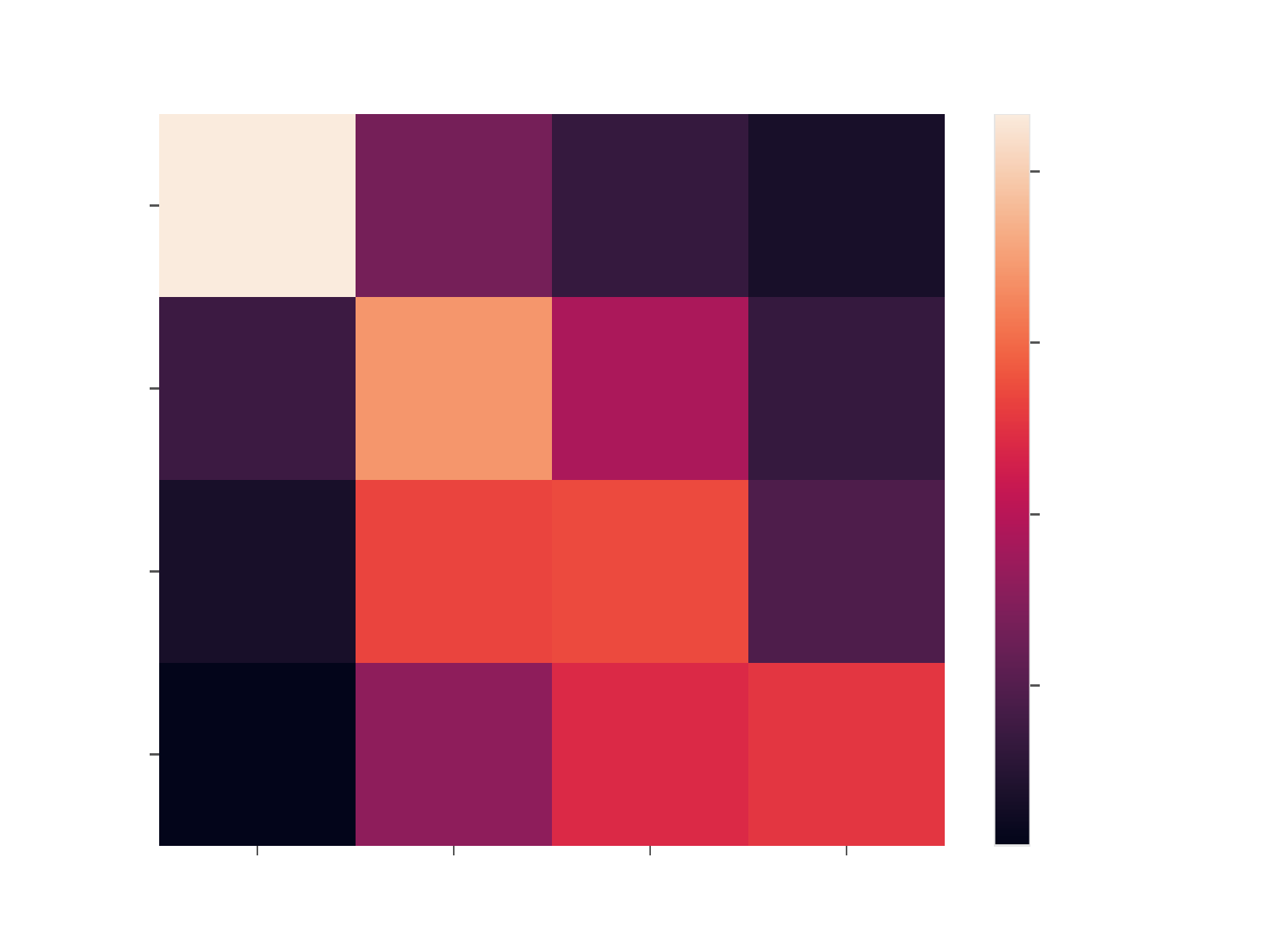}}%
    \put(0.17022149,0.57590114){\makebox(0,0)[lt]{\lineheight{1.25}\smash{\begin{tabular}[t]{l}0.65\end{tabular}}}}%
    \put(0.33000762,0.57590051){\color[rgb]{1,1,1}\makebox(0,0)[lt]{\lineheight{1.25}\smash{\begin{tabular}[t]{l}0.20\end{tabular}}}}%
    \put(0.48732387,0.57590114){\color[rgb]{1,1,1}\makebox(0,0)[lt]{\lineheight{1.25}\smash{\begin{tabular}[t]{l}0.10\end{tabular}}}}%
    \put(0.63716532,0.57590114){\color[rgb]{1,1,1}\makebox(0,0)[lt]{\lineheight{1.25}\smash{\begin{tabular}[t]{l}0.05\end{tabular}}}}%
    \put(0.33000825,0.43809073){\color[rgb]{1,1,1}\makebox(0,0)[lt]{\lineheight{1.25}\smash{\begin{tabular}[t]{l}0.51\end{tabular}}}}%
    \put(0.487324,0.43809136){\color[rgb]{1,1,1}\makebox(0,0)[lt]{\lineheight{1.25}\smash{\begin{tabular}[t]{l}0.28\end{tabular}}}}%
    \put(0.63716578,0.43809136){\color[rgb]{1,1,1}\makebox(0,0)[lt]{\lineheight{1.25}\smash{\begin{tabular}[t]{l}0.10\end{tabular}}}}%
    \put(0.1702218,0.43809056){\color[rgb]{1,1,1}\makebox(0,0)[lt]{\lineheight{1.25}\smash{\begin{tabular}[t]{l}0.11\end{tabular}}}}%
    \put(0.33000825,0.29108016){\color[rgb]{1,1,1}\makebox(0,0)[lt]{\lineheight{1.25}\smash{\begin{tabular}[t]{l}0.40\end{tabular}}}}%
    \put(0.48732397,0.29108079){\color[rgb]{1,1,1}\makebox(0,0)[lt]{\lineheight{1.25}\smash{\begin{tabular}[t]{l}0.41\end{tabular}}}}%
    \put(0.63716578,0.29108079){\color[rgb]{1,1,1}\makebox(0,0)[lt]{\lineheight{1.25}\smash{\begin{tabular}[t]{l}0.14\end{tabular}}}}%
    \put(0.1702218,0.29107996){\color[rgb]{1,1,1}\makebox(0,0)[lt]{\lineheight{1.25}\smash{\begin{tabular}[t]{l}0.05\end{tabular}}}}%
    \put(0.33000828,0.14704469){\color[rgb]{1,1,1}\makebox(0,0)[lt]{\lineheight{1.25}\smash{\begin{tabular}[t]{l}0.24\end{tabular}}}}%
    \put(0.487324,0.14704536){\color[rgb]{1,1,1}\makebox(0,0)[lt]{\lineheight{1.25}\smash{\begin{tabular}[t]{l}0.36\end{tabular}}}}%
    \put(0.63716578,0.14704536){\color[rgb]{1,1,1}\makebox(0,0)[lt]{\lineheight{1.25}\smash{\begin{tabular}[t]{l}0.38\end{tabular}}}}%
    \put(0.1702218,0.14704449){\color[rgb]{1,1,1}\makebox(0,0)[lt]{\lineheight{1.25}\smash{\begin{tabular}[t]{l}0.01\end{tabular}}}}%

    \put(0.15255453,0.04235919){\makebox(0,0)[lt]{\lineheight{1.25}\smash{\begin{tabular}[t]{l}Healthy\end{tabular}}}}%
    \put(0.32188649,0.04235919){\makebox(0,0)[lt]{\lineheight{1.25}\smash{\begin{tabular}[t]{l}Mild\end{tabular}}}}%
    \put(0.4433799,0.04235919){\makebox(0,0)[lt]{\lineheight{1.25}\smash{\begin{tabular}[t]{l}Moderate\end{tabular}}}}%
    \put(0.61994623,0.04235919){\makebox(0,0)[lt]{\lineheight{1.25}\smash{\begin{tabular}[t]{l}Severe\end{tabular}}}}%

    \put(0.10560577,0.53534556){\rotatebox{90}{\makebox(0,0)[lt]{\lineheight{1.25}\smash{\begin{tabular}[t]{l}Healthy\end{tabular}}}}}%
    \put(0.10450317,0.40847031){\rotatebox{90}{\makebox(0,0)[lt]{\lineheight{1.25}\smash{\begin{tabular}[t]{l}Mild\end{tabular}}}}}%
    \put(0.107034,0.23582015){\rotatebox{90}{\makebox(0,0)[lt]{\lineheight{1.25}\smash{\begin{tabular}[t]{l}Moderate\end{tabular}}}}}%
    \put(0.10391213,0.10920569){\rotatebox{90}{\makebox(0,0)[lt]{\lineheight{1.25}\smash{\begin{tabular}[t]{l}Severe\end{tabular}}}}}%

    \put(0.82773665,0.19944044){\makebox(0,0)[lt]{\lineheight{1.25}\smash{\begin{tabular}[t]{l}0.15\end{tabular}}}}%
    \put(0.82773698,0.33387988){\makebox(0,0)[lt]{\lineheight{1.25}\smash{\begin{tabular}[t]{l}0.30\end{tabular}}}}%
    \put(0.82773698,0.47059833){\makebox(0,0)[lt]{\lineheight{1.25}\smash{\begin{tabular}[t]{l}0.45\end{tabular}}}}%
    \put(0.82773698,0.60443711){\makebox(0,0)[lt]{\lineheight{1.25}\smash{\begin{tabular}[t]{l}0.60\end{tabular}}}}%
  \end{picture}%
\endgroup%